# A Sheath Collision Model with Thermionic Electron Emission and the Schottky Correction Factor for Work Function of Wall Material

L. Pekker

Victor Technologies, West Lebanon, NH 03784

## Abstract

This paper proposes a model that expands Godyak's collisional sheath model to the case of hot electrodes (anode or cathode) with thermionic electron emission. In the model, the electrodes are assumed to be made from refractory metals and, consequently, the erosion of the electrodes is small and can be neglected. In the frame of two temperature thermal plasma modeling, this model allows self-consistent calculation of the sheath potential drop, the Schottky correction factor for the work function of the wall material, the thermionic electron current density, and the heat fluxes of the charged particles from the plasma to the wall. The model is applied to the cathode spot at the tungsten cathode in argon. It is shown that the Schottky correction factor plays a crucial role in modeling high-intensity arcs. It is demonstrated that a virtual cathode can be formed in the atmospheric pressure argon plasma at the cathode surface temperature of 4785 K if the cathode current density is sufficiently small. The heat flux to the thermionic cathode due to charged particles and the heat flux to the plasma due to thermionic electrons are calculated. The model can be reduced to the case of the cold walls where the thermionic electron emission and the wall erosion processes are small and can be neglected. The sheath potential drop and the heat fluxes calculated by this model can be used as boundary conditions at the wall for the electric potential and for the energy equations for the electrons and heavy particles (ions and neutrals).

## I. Introduction

The formation of plasma sheath at the wall plays a fundamental role in the heat flux from the plasma to the wall, thermionic electron emission, the structure of the cathode spot and anode attachment, the electrode erosion process, and other electrode process. Therefore, the plasma sheath has to be taken into account in modeling high-pressure thermal arcs (i.e., whenever the plasma pressure is as large as or larger than atmospheric pressure). In recent works [1, 2] the authors constructed boundary conditions at the wall for the electric potential and the electron and heavy particle energy equations in two temperature, $T_e \neq T_h$, modeling of "thermal" plasmas; $T_e$ is the temperature of electrons and $T_h$ is the temperature of heavy particles, ions and neutrals. In these works, the sheath is viewed as the interface between the plasma and the wall. The case of a cold wall, where the thermionic electron emission and the erosion of the wall are small and can be neglected, was considered in [1], and the case of a hot wall with thermionic electron emission was considered in [2]. In [2], the wall is assumed to be made from refractory metals that the erosion of the wall is small and can be neglected. In [1], the authors used the Godyak collision sheath model [3, 4], and in [2] they expanded it to the case of the electron thermionic emission. In both works [1] and [2], the Schottky effect describing the decrease in the work function of the material in a strong electric field was calculated by neglecting the friction of ions in the sheath. Thus, in the case of a collisional sheath, calculations of the thermionic electron current density and the heat flux to the wall in these models are not self-consistent and, therefore, are inaccurate.

In the present paper, I construct a collisional sheath model that takes into account the friction of ions in the sheath in calculating the Schottky correction factor. This allows one to calculate the thermionic electron emission and the heat flux to the wall in a consistent way. The collisional sheath model with thermionic electron emission is presented in Section II. In Section III, I applied the model to the case of the of a tungsten cathode in argon plasma with parameters that are typical for plasma cutting arcs, where the cathode surface temperature is relatively small and the virtual cathode is not formed. The obtained results are compared with model [2], where the Schottky correction factor was calculated by neglecting the friction of ions in the sheath. The case of an extremely high cathode surface temperature where the virtual cathode is formed is considered in Section IV. The conclusions are given in Section VI.

## II. Collision sheath model and heat fluxes with thermionic electron emission

As in [1, 2], the following assumption are made in the model: (1) the plasma is considered in two-temperature thermal approximation, $T_e \neq T_h$; (2) the chemical plasma equilibrium is achieved; (3) the plasma sheath is viewed as an interface between the plasma and the wall; (4) the plasma at the wall is singly ionized; (5) the potential in the sheath is monotonically decreasing from the plasma side to the wall; (6) $T_e \gg T_h = T_{sur}$, where $T_{sur}$ is the temperature of the surface of the wall; (7) the friction of ions with electrons is neglected relative to the friction of ions with neutrals; and (8) thermionic electrons pass through the sheath collisionlessly, transferring their momentum and energy far from the sheath in the plasma; in other words, the thickness of the sheath is much smaller than the transport mean free path for thermionic electrons:

$$L_{sheath} \ll \lambda_{e-therm-mfp}. \qquad (1)$$

Following [2], a general formula for the sheath potential drop at the cathode with thermionic emission can be written as

$$\varphi_{sheath} = -\frac{k_B T_e}{e} \ln\left[\sqrt{\frac{2\pi m_e}{M\left(1+\frac{\pi r_{De}}{2\lambda_{i-mfp}}\right)}}\left(1 - \frac{j-j_{e-therm}}{j_i}\right)\right], \qquad (2)$$

where $k_B$ is the Boltzmann constant; $e$ is the charge of an electron; $r_{De} = \left(\varepsilon_0 k_B T_e/n_p e^2\right)^{1/2}$ is the electron Debye radius; $j$ is the total cathode current density; $\lambda_{i-mfp} = 1/n_n \sigma_{i,n}$ is the ion transport mean free path; $n_n$ is the number density of neutral atoms; $\sigma_{i,n}$ is the charge-exchange cross section that is independent of the ion velocity, the dominant ion-neutral momentum transfer process in the sheath [3, 4];

$$j_{e-plasma} = en_p \exp\left(-\frac{e\varphi_{sheath}}{k_B T_e}\right)\sqrt{\frac{k_B T_e}{2\pi m_e}} \qquad (3)$$

is the electron plasma current density in the sheath; $n_p$ is the plasma number density (in the case of singly ionized plasma $n_p = n_i = n_e$, where $n_i$ and $n_e$ are the electron and ion plasma number densities respectively);

$$j_i = en_p V_s = en_p \sqrt{\frac{k_B T_e}{M\left(1+\frac{\pi r_{De}}{2\lambda_{i-mfp}}\right)}} \qquad (4)$$

is the ion current in the sheath; $V_s$ is the ion velocity at which the ions enter the sheath;

$$j_{e-therm} = AT_{sur}^2 \exp\left(-\frac{e(\varphi_{work-func} - \Delta\varphi_{Schot})}{k_B T_{sur}}\right) \qquad (5)$$

is the electron thermionic current density in the sheath; and

$$j = j_{e-therm} + j_i - j_{e-plasma} \quad . \qquad (6)$$

In Eq. (5), $\varphi_{work-func}$ is the work function of the cathode material; $\Delta\varphi_{Schot}$ is the Schottky correction factor,

$$\Delta\varphi_{Schot} = \left(-\frac{eE_{sur}}{4\pi\varepsilon_0}\right)^{1/2}, \tag{7}$$

describing the decrease of the effective work function of materials in strong electric fields; $E_{sur}$ is the electric field at the surface of the wall; and $A$ depends on the cathode material. The thickness of the cathode sheath is on the order of $r_{De}$.

Let us introduce a collision factor as

$$\alpha_{col} = \frac{\pi r_{De}}{2\lambda_{i-mfp}}. \tag{8}$$

The case of $\alpha_{col} \ll 1$ corresponds to the collisionless Bohm's sheath [5], where the ions are freely accelerated in the sheath, and the case $\alpha_{col} > 1$ corresponds to the collisional sheath [3, 4], where the ions move in the sheath in the charge exchange regime. In the case of very high gas pressure, $\alpha_{col} \gg 1$ and $V_s < \sqrt{k_B T_h/M}$, the ions move in the sheath in the mobility (not charge exchange) regime ($\lambda_{i-mfp}$ in the mobility regime is dependent on the ion velocity), and Eq. (2) should be modified accordingly [4]; this case is not considered in the present paper.

To obtain $j_{e-therm}$, Eq. (6), at given plasma and wall parameters: $j$, $n_p$, $n_n$, $T_e$ and $T_h = T_{sur}$ I have to calculate $E_{sur}$ by solving the Poisson equation in the cathode sheath. An equation for ion movement in the plasma sheath can be written as:

$$MV_i \frac{dV_i}{dx} - e\frac{d\varphi}{dx} + F_{i-frick} = 0, \tag{9}$$

where $V_i$ is the velocity of an ion in the sheath,

$$F_{i-frick} = \pi \frac{MV_i^2}{2\lambda_{i-mfp}} \tag{10}$$

is the friction force of ions with neutrals [4], $x$ is the coordinate axis directed from the plasma to the wall, Fig. 1, and $\varphi$ is the electric potential in sheath; $x$ and $\varphi$ are equal to zero at the boundary between the plasma with the sheath, Fig. 1, and

$$V_i(x=0) = V_s = \sqrt{\frac{k_B T_e}{M\left(1 + \frac{\pi r_{De}}{2\lambda_{i-mfp}}\right)}}. \tag{11}$$

Taking into account that the ion current is conserved in the sheath (no ionization and recombination in the sheath), $j_i = en_i V_i$, the Poison equation can be written as:

$$\frac{d^2\varphi}{dx^2} = \frac{e}{\varepsilon_0}\left(n_p exp\left(\frac{e\varphi}{k_B T_e}\right) - \frac{j_i}{eV_i} + \frac{j_{e-therm}}{e\sqrt{\frac{2e(\varphi_{sheath}+\varphi)}{m_e}}}\right), \quad (12)$$

where $j_i$ is given by Eq. (4). In Eq. (12), the first term in the parentheses is the density of "plasma" electrons, the second term is the density of ions, and the third term is the density of "thermionic" electrons in the sheath. It should be stressed that in [2], the friction force $F_{i-frick}$ in Eq. (9) was omitted -- that is literally correct only in the case of collisionless sheath. As in [2-4], the boundary conditions for Eq. (12) are

$$\varphi(x=0) = 0, \quad \left(\frac{d\varphi}{dx}\right)_{x=0} = -\frac{k_B T_e}{er_{De}}, \quad \varphi(L_{sheath}) = -\varphi_{sheath}, \quad (13)$$

where the first condition states that the potential at the sheath from the plasma side is equal to zero, Fig. 1; the second condition is chosen according to the Godyak sheath model [3, 4]; and the third condition determines $L_0$, the length of the sheath, Fig. 1. Although, Godyak derived his boundary conditions for the case of no secondary electron emission, these boundary conditions can be also applied for thermionic electrodes. As has been mentioned in [4], the second condition, in fact, describes the "electrostatic wall" separating electrons from the wall. This is reasonable because the density of plasma electrons at the cathode in the model is assumed to be much smaller than in the plasma, $exp(-e\varphi_{sheath}/k_B T_e) \ll 1$.

As follows from Eq. (9), the plasma at $x = 0$ is not quasineutral, and the charge density at $x = 0$ is

$$\Delta\rho(x=0) = \frac{j_{e-therm}}{\sqrt{\frac{2e\varphi_{sheath}}{m_e}}}. \quad (14)$$

As a result, $\varphi_{cath}$ calculated by the model is smaller than the "real" potential drop between the quasineutral plasma and the wall; this difference $\Delta\varphi$ can be estimated as

$$\Delta\varphi = -\frac{k_B T_e}{e} ln\left(1 - \frac{j_{e-therm}}{en_p\sqrt{\frac{2e\varphi_{sheath}}{m_e}}}\right). \quad (15)$$

Thus, the suggested model is reasonable only if the condition,

$$\frac{\Delta\varphi}{\varphi_{sheath}} = -\frac{k_B T_e}{e\varphi_{sheath}} \ln\left(1 - \frac{j_{e-therm}}{en_p\sqrt{\frac{2e\varphi_{sheath}}{m_e}}}\right) \ll 1, \qquad (16)$$

is well satisfied [2].

Integration of the set of Eqs. (9) and (12) with the boundary conditions given by Eqs. (11) and (13) yields $\varphi(x)$ and $V_i(x)$ in the sheath, and also $L_{sheath}, E_{sur}, \Delta\varphi_{Schot}, \varphi_{sheath}$, and $j_{e-therm}$.

However, at sufficiently high $j_{e-therm}$ this system of equations may not have a real solution. This corresponds to the situation where the electric potential distribution in the sheath is no longer a monotonic function of $x$ as the model assumed, assumption (4), and a "virtual cathode" cathode is formed at the cathode surface [2], Fig. 2. In the case of a virtual cathode, not all electrons emitted from the cathode reach the plasma, and some of them are rebounded back into the cathode, leading to a decrease in the actual thermionic electron current passing through the sheath. So, it is necessary to determine the conditions of the formation of the virtual cathode and calculate the thermionic electron emission current density where the vertical cathode is formed. The following procedure was suggested in [2]: (1) Solve the set of Eqs. (9) and (12) with boundary conditions given by Eqs. (11) and (13) at given plasma parameters: $j, n_p, n_n, T_e$ to determine the critical value of the thermionic current density, $j_{e-therm}^{critical}$ and $\varphi_{sheath}^{critical}$, which correspond to $E_{sur} = 0$; (2) Calculate the thermionic current density at given $T_{sur}$ and $\Delta\varphi_{Schot} = 0$; (3) If the resulting $j_{e-therm}$ is smaller than $j_{e-therm}^{critical}$, then the virtual cathode is not formed and the system of Eqs. (9) and (12) with boundary conditions given by Eq. (11) and (13) is solvable. If not and $j_{e-therm}^{critical} < j_{e-therm}$, then $j_{e-therm}^{critical}$ and $\varphi_{sheath}^{critical}$ have to be used. It is worth noting that in the case of the virtual cathode, the effective length of the cathode sheath is the distance between the plasma-sheath interface and the "emitting-virtual-cathode surface" $L_{sheath}^{critical}$, Fig. 2.

In addition, it is worth noting that in the case of the thermionic anode, a formula for the anode sheath potential drop can be obtained from Eq. (2) by using $-j$ instead $j$ [2],

$$\varphi_{sheath} = -\frac{k_B T_e}{e} \ln\left[\sqrt{\frac{2\pi m_e}{M\left(1+\frac{\pi r_{De}}{2\lambda_{i-mfp}}\right)}} \left(1 + \frac{j+j_{e-therm}}{j_i}\right)\right], \qquad (17)$$

because in the case of anode, the current is directed from the wall to the plasma while in the case of the cathode, it is directed from the plasma to the wall. In the case of a cold cathode with no thermionic electron emission, $\varphi_{sheath}$ can be obtained from Eq. (2) by omitting $j_{e-therm}$, and in the case of a cold anode, by omitting $j_{e-therm}$ in Eq. (17). In the case of cold floating walls ($j = j_{e-therm} = 0$), $j_{e-therm}$ and $j$ have to be dropped in Eq. (2).

Following [1, 2] the enthalpy flux from the singly ionized plasma to the wall due to charged particles can be written as

$$q_{charged}^{particles} = q_{ions} + q_{e-plasma}, \tag{18}$$

$$q_{ions} = en_p V_s \left(I_{ioniz} + \varphi_{sheath} + \frac{MV_s^2}{2e}\right), \tag{19}$$

$$q_{e-plasma} = 2k_B T_e n_p \exp\left(-\frac{e\varphi_{sheath}}{k_B T_e}\right) \sqrt{\frac{k_B T_e}{2\pi m_e}}, \tag{20}$$

where $q_{ions}$ is the ion heat flux from the plasma to the wall, which includes the heat flux to the wall due to the recombination process plus the kinetic energy flux that ions brings to the wall (directly, or by fast atoms created in the charge exchange process); $q_{e-plasma}$ is the heat flux that plasma electrons bring to the wall; $I_{ioniz}$ is the ionization potential of the working gas; $MV_s^2/2$ is the kinetic energy of an ion entering the sheath; $V_s$ is given by Eq. (4); and $\varphi_{sheath}$ is given by Eq. (2), the case of no virtual cathode, or by $\varphi_{sheath}^{critical}$, the case of a virtual cathode. Eq. (19) assumes that all ions entering the sheath reach the wall, recombine there with electrons, and come back to the plasma as neutrals, where they are immediately ionized by electrons. Since the model assumes that $k_B T_h \ll e\varphi_{anode}$, the ion thermal heat flux to the wall is neglected in Eq. (19).

Taking into account the energy flux that the cathode loses due to the "condensation" energy of electrons at the wall, the total heat flux to the cathode due to charged particles can be written as [2],

$$Q_{charged}^{particles} = q_{charged}^{particles} + q_{cond}, \qquad q_{cond} = -j(\varphi_{work-func} - \Delta\varphi_{Schot}), \tag{21}$$

where $\varphi_{work-func}$ is the work function of the anode material and $\Delta\varphi_{Schot}$, Eq. (7), is determined by solving the set of Eqs. (9) – (13). Since the current is directed to the cathode the condensation heat flux in Eq. (21) is negative. Thus, at a given $n_p$, $n_n$, $T_e$, and $j$, the heat flux to the wall due to charged particles $Q_{charged}^{particles}$, Eq. (21), can be calculated. In the case where the wall is an anode, $Q_{charged}^{particles}$ can be obtained from Eq. (21)

by changing the minus sign in front of $j$ to a plus sign; and in the case of the floating wall ($j = 0$), by dropping $q_{cond}$ in Eq. (21) entirely.

The total heat flux to the wall due to all particles, neutral and charged, can written as,

$$Q_{total}^{particles} = -\kappa_n \frac{\partial T_h}{\partial n} + Q_{charged}^{particles} \qquad (22)$$

where $\partial T/\partial n$ is the space derivative of $T$ normal to the wall and the first term in the right-hand side of Eq. (22) is the heat flux of the neutral gas molecules to the wall. In Section III, I analyze the heat flux due to charged particles in the case of the cathode spot formed at a thermionic tungsten cathode in argon plasma. The thermal heat flux of the neutral particles to the wall cannot be calculated in the frame of the present model and therefore is not considered in here. According to [6], in the case of a free burning arc in argon, the heat flux contribution from neutrals to the cold anode can range from 20% to 60% depending on the anode current density and geometry.

The heat flux that the thermionic electrons bring to the plasma, $q_{e-therm}$, is

$$q_{e-therm} = j_{e-therm} \cdot \varphi_{sheath}. \qquad (23)$$

In Eq. (23) I have neglected the thermal energy of the thermionic electrons because $k_B T_{sur} \ll e\varphi_{sheath}$. Thus, the enthalpy flux that the plasma is losing at the wall due to charged particles is $Q_{i+e}^{plasma} = q_{charged}^{particles} - q_{e-therm}$.

It has to be stressed that the $\varphi_{sheath}$ and $j_{e-therm}$ (or $\varphi_{sheath}^{critical}$ and $j_{e-therm}^{critical}$ in the virtual electrode case) obtained in the model can be used in the boundary conditions for the electric potential, while the heat fluxes in the boundary conditions for the energy equations for electrons and heavy particle in the way described in [2].

### III. Thermionic cathode: numerical results

In this section, I illustrate the sheath model applied to the case of the thermionic electron emission at a tungsten cathode in a singly ionized argon plasma at given $T_e$, $T_{sur}$, and $P$. Following [7], plasma composition in this case can be determined by solving the Saha equation:

$$\frac{n_p^2}{n_n} = 2\left(\frac{2\pi m_e k_B T_e}{h^2}\right)^{3/2} \frac{Q_{Ar^+}(T_e)}{Q_{Ar}(T_e)} exp\left(-\frac{eI_{ioniz}}{k_B T_e}\right) = 2.89 \times 10^{22} T_e^{3/2} exp\left(-\frac{1.827 \times 10^5}{T_e}\right), \qquad (24)$$

$$P = k_B(n_p + n_n)T_h + k_B n_e T_e , \qquad (25)$$

where $Q_{Ar^+}(T_e)$ and $Q_{Ar}(T_e)$ are the statistical sums of partition functions of argon ions and argon neutral atoms respectively. Two assumptions were made in Eq. (24) and (25): (1) the contributions of the excited states to the statistical sums $Q_{Ar^+}$ and $Q_{Ar}$ are neglected in Eq. (25) since they are known to be less than 5 percent [8]; and (2) multi-charged ions are ignored in this model because the number densities of multi-charged ions are many orders of magnitude smaller than the number density of singly ionized argon.

In this Section, I consider the case of a tungsten emitter at a relatively small surface temperature of 3800K, $P = 4 \cdot 10^5$Pa, and $T_e = 9000$K; these parameters are typical for plasma cutting torches. Solving Eqs. (17) and (18) with $T_h = T_{sur}$ yields $n_n = 7.57 \cdot 10^{24} m^{-3}$ and $n_p = 1.69 \cdot 10^{22} m^{-3}$. Using the obtained plasma composition, $\sigma_{i,n} = 1.18 \cdot 10^{-18} m^2$ [9] (the total $A^+ - Ar$ momentum transfer cross-section), $A = 6 \cdot 10^5$ A/(m²·K²) and $\varphi_{work-func} = 4.54$ eV (the Richardson parameters of tungsten), I obtain that: $\lambda_{i-mfp} = 1.11 \cdot 10^{-7} m$; $r_{De} = 5.05 \cdot 10^{-8} m$; $\alpha_{col} = 0.7$, Eq. (8); $V_s = 1046 m/sec$, Eq. (4), $j_{e-therm}(\Delta\varphi_{Schot} = 0) = 8.2 \cdot 10^6 A/m^2$, Eq. (5); $j_i = 2.82 \cdot 10^6 A/m^2$, Eq. (4). The results of the simulation are presented in Figs. 3 – 8. In Figs. 3 – 6, I compare the obtained results with the results obtained by model [2] in which the third term in the left-hand side of Eq. (9) is dropped.

It should be stressed that this simulation cannot be considered as a cathode spot model because the model does not consider the total heat balance between the plasma and the cathode. The model does not calculate various important quantities: (1) the conduction heat flux from the plasma to the wall due to neutral particles, (2) the radiation cooling of the cathode, (3) the electron convective heat flux, (4) the heat transfer in the cathode, (5) the Ohmic heating of the cathode, and other heat transfer processes which must be included in the total heat balance between the plasma and the wall, see for example [2, 11] and references therein. The purpose of this simulation is to illustrate that taking into account the friction of ions in the sheath while calculating the Schottky correction factor is important for modeling collisional sheaths at thermionic electrodes.

As one can see from Figs. 3 – 5 the sheath potential drop, the Schottky decrease in the work function of the tungsten cathode, and the length of the cathode sheath increase with an increase in the total cathode current density $j$, as was expected; $j = 0$ corresponds to the case of a floating wall. As one can see from Fig. 3, even if the total cathode current density is equal to zero, the sheath is still formed at the thermionic electrode to block the plasma electrons from reaching the wall. As follows from Fig. 4, $\Delta\varphi_{Schot}$ is highly dependent on the total current density and dramatically affects the work function of the material; as shown in Fig. 6, $j_{e-therm}$ reaches its maximum value of $1.81 \cdot 10^6 A/m^2$ which is 2.2 times larger than $j_{e-therm}$ at $\Delta\varphi_{Schot} = 0$.

As follows from Fig. 6 the current density of plasma electrons sharply decreases with an increase in $j$ and becomes negligibly small for $j > 2 \cdot 10^7 A/m^2$; this was expected because the sheath potential drop increases with $j$, Fig. 3; consequently, this leads to the sharp decrease in $j_{e-plasma}$, Eq. (3); $j_{e-plasma}$ at $\varphi_{sheath} = 0$ (no sheath) is $3.98 \cdot 10^8 A/m^2$. Thus, for $j > 2 \cdot 10^7 A/m^2$, $j$ consists, essentially, of $j_i$ and $j_{e-therm}$, Eq. (6). It should be stressed that the system of Eqs. (9) and (12) has no solution for $j > 2.1 \cdot 10^6 A/m^2$ because the $j_i$ is a constant independent of $j$, Eq. (4), and $j_{e-therm}$ is limited because the Schottky correction factor reaches its maximum of 0.2601 eV as $\varphi_{sheath} \to \infty$ which corresponds to $j = j_{max} = 2.1 \cdot 10^6 A/m^2$. This observation makes perfect sense: the thermionic cathode simply cannot provide enough thermionic electrons to support larger current densities.

As one can see from Figs. 3 – 6, the present model and model [2] give notably different results. Since model [2] neglects the friction of ions in Eq. (9), the third term in the left-hand side in this equation is absent in model [2], the electric field required to maintain the ion current density in the sheath, Eq. (4), calculated by model [2] has to be smaller than that calculated by the present model. Therefore, at given $j$: (1) the Schottky correction factor, Eq. (7), and the thermionic electron current density, Eq. (5), calculated by model [2] are smaller than those calculated by present model, Figs. 4 and 6; and (2) the sheath length calculated by model [2] is larger than that calculated by the present model. Because at a given cathode current density, $j$, $j_{e-therm}$ calculated by present model is larger than that calculated by model [2], the sheath potential drop calculated by model [2] is always larger than that calculated by the present model, Fig. 3. This leads to smaller plasma electron current densities calculated by model [2] than those calculated by presented model, Fig. 6. The cathode length calculated by model [2] increases more rapidly than the cathode length calculated by present model, Fig. 3, because the $j_{max}$ calculated by model [2] is $1.96 \cdot 10^6 A/m^2$ and smaller than $j_{max}$ calculated by the present model.

The heat fluxes of the charged particles at the plasma-cathode interface are shown in Figs. 7 and 8. As one can see from Fig. 7, because the sheath potential drop, $\varphi_{sheath}$, increases with the total cathode current density, Fig.3, $q_{e-plasma}$ decreases and $q_{ion}$ increases with an increase in $j$, Eqs. (19) and (20). Because for the all current densities considered, $\Delta\varphi_{Schot}/\varphi_{work-func} \ll 1$, Fig. 4, $q_{cond}$, Eq. (21), decreases almost linearly with an increase in $j$, Fig. 7. As follows from Fig. 7, for $j > 1.4 \cdot 10^7 A/m^2$ the cathode is losing more energy due to "vaporizing" electrons, $q_{cond}$, than it gains from the heat flux from the plasma to the wall due charged particles. As follows from Fig. 8, for $j > 1.6 \cdot 10^7 A/m^2$ the plasma gains more energy from the thermionic electrons accelerated in the sheath than it loses due to heat fluxes from the plasma to the wall due to the charged particles.

Plasma sheath model [3, 4] assumes that $E_{plasma}$, the electric field in the plasma at the plasma-sheath interface, has to be much smaller than $k_B T_e/er_{De}$, the electric field in the sheath at the sheath side, Eq. (13).

Now I will demonstrate that this assumption is valid for this simulation. Following [2], substituting in the plasma conductivity formula $\sigma = e^2 n_p/(v_{e,i}^{tr} + v_{e,n}^{tr})$ the values of $v_{e,i}^{tr}$ and $v_{e,n}^{tr}$ (the electron-ion and electron-neutral transport collision frequencies),

$$v_{e,i}^{tr} = 4.93 \cdot 10^{-6} \frac{\Lambda n_p}{T_e^{3/2}}, \quad \Lambda = 18.7 - \ln\left(\frac{n_p^{1/2}}{T_e^{5/4}}\right), \quad v_{e,n}^{tr} = n_n \sigma_{e,n} \sqrt{\frac{k_B T_e}{m_e}}, \quad \sigma_{e,n} = 2 \cdot 10^{-20} m^2, \quad (26)$$

yields $\sigma = 9.7 \cdot 10^2$ S/m. In Eq. (26), $\sigma_{e,n}$ was extracted from the data in [10]. Since in the model, the electric field in the sheath at the plasma-sheath interfaces is independent of the cathode current density and the electric field in the plasma at the plasma-sheath interface, $E_{plasma} = j/\sigma$, increases with an increase in the cathode current density, I obtain that $E_{plasma}$ reaches its maximum of $22.2 \cdot 10^4$ at $j = 2.1 \cdot 10^6 A/m^2$. This is 711 times smaller than $k_B T_e / e r_{De}$, justifying the assumption.

Now I validate the model assumption that in the sheath, the ion-electron friction force is much smaller than the ion-neutral friction force. The ratio of the ion-electron friction force to the ion-neutral friction force in the sheath can be estimated as [2]:

$$Friction_{i,n}^{i,e} \sim \frac{m_e \cdot (v_{i,e-plasma}^{tr} + v_{i,e-therm}^{tr})}{M v_{i,n}^{tr}}, \quad (27)$$

where $v_{i,e-plasma}^{tr}$ and $v_{i,e-therm}^{tr}$ are the collision frequencies of an ion with plasma electrons and thermionic electrons respectively, and $v_{i,n}^{tr} \approx n_n \sigma_{i,n} V_s$. Substituting $j_{e-therm}/e\sqrt{2e\varphi_{sheath}/m_e}$ and $e\varphi_{sheath}$, the characteristic thermionic electron number density and the characteristic energy of thermionic electons in the sheath, into Eq. (26) instead of $n_p$ and $k_B T_e$, one can estimate $v_{i,e-therm}^{tr}$. Substituting the number density $n_p = 1.69 \cdot 10^{22} m^3$ and the temperature of the plasma electrons in the sheath $T_e = 9000$ K into Eq. (26), I obtain an upper estimate for $v_{i,e-plasma}^{tr}$ in the sheath; in the sheath, the number density of plasma electrons is maximal at the plasma-sheath interface. For the range of $j_{e-therm}$ considered in this simulation, $Friction_{i,n}^{i,e}$ is smaller than $6.3 \times 10^{-4}$. Thus, neglecting the friction of ions with electrons in this simulation is appropriate.

Now I review the validity of the assumption that thermionic electrons pass through the sheath collisionlessly. Following [2], substituing $n_p = 1.69 \cdot 10^{22} m^{-3}$ (the number density of plasma electrons at the plasma-sheath interface) and $\Lambda = 5$ (the characteristic Coloumb logarithm in the sheath) in

$$\frac{1}{\lambda_{e-therm-mfp}} = n_p \frac{6.43 \cdot 10^{-10} \Lambda}{\left(\frac{e \cdot \varphi_{sheath}}{k_B}\right)^2} + n_n \sigma_{e,n}, \qquad (28)$$

I obtain that the ratio of $\lambda_{e-therm-mfp}$ to $L_{sheath}$ is greater than 23 in the full range of $j$ considered in this simulation. The first term in the right-hand side in Eq. (21) describes the collisions of a thermionic electron with plasma electrons and ions, and the second term describes the collisions with neutrals. Thus, neglecting the collisions of the thermionic electrons in the sheath is appropriet in this simulation.

Next, I validate the model assumption that in the sheath ions move in the charge-exchange regime. Because the ions enter the sheath with velocity $V_s = 1046 m/sec$ (see above) which is larger than $\sqrt{k_B T_h/M} = 889 m/sec$, the ions indeed move in the sheath in the charge-exchange regime [4].

Now let us check that $\Delta\varphi/\varphi_{sheath} \ll 1$, Eq. (16). Substituting the calculated functions of $\varphi_0$ and $j_{e-therm}$ vs. $j$ into Eq. (16) I obtain that that $\Delta\varphi/\varphi_{sheath} < 0.002$ in the entire range of $j$ considered in the simulation. Thus, neglecting $\Delta\varphi$ in the model is appropriate.

In summary, I validated the model assumption that the evaporation of the cathode material is small and can be neglected. Substituting $H = 774 kJ/mol$, the tungsten heat of vaporization, $T_{boil} = 6203 K$, the boiling point of Tungsten, and $T_{sur} = 3800 K$ in the Clausius-Clapeyron equation,

$$P_{equl-tungsten} = 10^5 \cdot exp\left(\frac{H}{R} \cdot \left(\frac{1}{T_{boil}} - \frac{1}{T_{sur}}\right)\right), \qquad (29)$$

I obtain that the equilibrium partial pressure of the tungsten vapor is 7.56 Pa, which is negligibly small compared to the pressure of the argon plasma.

Let us next estimate the tungsten plasma density assuming that all evaporated tungsten atoms are ionized and come back to the cathode as ions, in other words, there is no erosion of the tungsten cathode at all. Setting the flux of the tungsten atoms that left the cathode equal to the flux of the tungsten ions that reached the cathode I obtain the following equation,

$$\frac{P_{equl-tungsten}}{k_B T_{sur}} \cdot \sqrt{\frac{k_B T_{sur}}{2\pi M}} = n_{p-tungsten} \sqrt{\frac{k_B T_e}{M}}. \qquad (30)$$

Solving Eq. (30) for $n_{p-tungsten}$ leads to an upper bound on the tungsten plasma number density at the plasma-sheath interface. In Eq. (30), I have used the Bohm velocity of the ions in the sheath [5]. Substituting $P_{equl-tungsten}$ from Eq. (29) into Eq. (30) yields $n_{p-tungsten} = 3.74 \cdot 10^{19} m^{-3}$ which is much smaller than the argon plasma density. Thus, neglecting the tungsten vapor in the model is appropriate.

It should be stressed that in this simulation the rate of thermionic electron emission is small (less than $j_{e-therm}^{critical}$) and, therefore, the virtual cathode is not formed at the cathode surface; even at $j = 0$, $\Delta\varphi_{Schot} > 0$, Fig. 4. As $j$ becomes negative, Fig. 3, the thermionic cathode becomes a thermionic anode. Although the case of a thermionic anode is not considered in the present paper, I would like to note that $\varphi_{sheath}$ continues decreasing with a decrease in $j$ in the case of a thermionic anode as well. At some values of the anode current, $\exp(-e\varphi_{sheath}/k_B T_e)$ approaches unity, and the model becomes invalid.

## IV. Virtual cathode: numerical results

In this Section, I consider the case where the thermionic electron emission current density is so large that the virtual cathode can be formed at the cathode. In this simulation I will take $T_{sur} = 4785K$, $P = 10^5 Pa$ and $T_e = 9000K$; the tungsten cathode at this temperature is molten. Solving Eqs. (24) and (25) with $T_h = T_{sur}$ yields $n_n = 1.49 \cdot 10^{24} m^{-3}$ and $n_p = 7.48 \cdot 10^{21} m^{-3}$. Using the obtained plasma composition I obtain that $\lambda_{i-mfp} = 5.68 \cdot 10^{-7}m$; $r_{De} = 7.58 \cdot 10^{-8}m$; $\alpha_{col} = 0.21$, $V_s = 1243 m/sec$, $j_{e-therm}(\Delta\varphi_{Schot} = 0) = 2.26 \cdot 10^8 A/m^2$; $j_i = 1.49 \cdot 10^6 A/m^2$, for details see Section III. The results of the simulation are presented in Figs. 9 –12.

As one can see from Figs. 9 – 12, as in Section III, $\varphi_{sheath}$, $\Delta\varphi_{Schot}$, $L_{sheath}$, and $j_{e-therm}$ increase and $j_{e-plasma}$ decreases with an increase in the total cathode current density $j$; $j_i$ is a constant independent of $j$, Eq. (4). However, there is a principal difference between the case of extremely large thermionic electron current densities, i.e., extremely large $T_{sur}$, where the virtual cathode is formed at the thermionic cathode, and the case of moderate $j_{e-therm}$ considered in Section III, where the virtual cathode is not formed. As one can see from Fig. 12A, for $j < 2.1 \cdot 10^8 A/m^2$ the thermionic current density is smaller than Richardson's current density at $\Delta\varphi_{Shot} = 0$. This means that for these cathode current densities the electric potential in the sheath is not monotonically decreasing as in Fig. 1, but has a "dip" as in Fig. 2 and, therefore, the effective thermionic current density, the effective sheath potential drop, and the effective length of the sheath are $j_{e-therm}^{critical}$, $\varphi_{sheath}^{critical}$, and $L_{sheath}^{critical}$, while $\Delta\varphi_{Schot}$ has no meaning, Section II. In Fig. 12, a further decrease in $j$ leads to such small cathode sheath voltage drops, Fig. 9, that $exp(-e\varphi_{sheath}/k_B T_e)$ approaches unity, and the model becomes invalid.

It should be stressed that in this simulation, as in Section III, the total cathode current density that can be extracted from the thermionic cathode is also limited. The Schottky correction factor reaches its maximum value of 0.138 eV as $\varphi_{sheath} \to \infty$ which corresponds to $j = j_{max} = 3.17 \cdot 10^8 A/m^2$.

In conclusion, I review the validity of the assumptions made in the model. For this simulation in all regions of the cathode current densities considered: (1) the ratio of the electric field in the sheath at the sheath-plasma interface, $k_B T_e/er_{De}$, to the electric field in the plasma at the plasma side of this interface, $E_{plasma}$, is larger than 30; (2) the value of $Friction_{i,n}^{i,e}$, Eq. (27), is smaller than 0.0072; (3) the value of $\lambda_{e-therm-mfp}$, is at least 173 times larger than $L_{sheath}$, Eq. (1); (4) $V_s = 1243 \, m/sec > \sqrt{k_B T_h / M} = 998 \, m/sec$ which means that in the sheath ions move in the charge-exchange regime; (5) $\Delta\varphi/\varphi_{sheath} = 0.123$, Eq. (16); (6) $P_{equl-tungsten}/P = 0.012$; (7) $n_{p-tungsten} = 5.16 \cdot 10^{21} m^{-3}$, Eq. (30). Thus, assumptions (1) – (6) are well satisfied in this simulation. Since $n_{p-tungsten}$ is an upper estimate for the tungsten plasma density and smaller than $n_p = 7.48 \cdot 10^{21} m^{-3}$, the argon plasma density, it is unlikely that taking into account the tungsten evaporation will notably change the results obtained in this section.

**V. Conclusions**

The paper extends Godyak's collision sheath model to the case of thermionic electron emission which allows a self-consistent calculation of the sheath potential drop, the Schottky correction factor, the thermionic electron current density, and the length of the sheath. The model assumes that the cathode is made from a refractory metal and, consequently, the erosion of the wall is small and can be neglected. Unlike models [1, 2], where the Schottky correction factor was calculated by neglecting the friction of ions with neutral particles in the sheath, the present model explicitly takes into account the collisions of ions with neutrals while calculating $\Delta\varphi_{Schot}$ and, therefore, is free of this inconsistency. The sheath model is also modified to the case of cold electrodes (anode or cathode) and cold floating walls with no thermionic electron emission and wall erosion that allows to self-consistently calculate $\Delta\varphi_{Schot}$ in the collisional sheath as well.

It was demonstrated that in calculating the Schottky correction factor, neglecting the friction between the ions and the neutral particles in the sheath may lead to significant undervalued magnitudes in the thermionic electron current densities and, consequently, incorrect simulation of the arc and the heat transfer between the plasma and the wall.

Two regimes of the arc at the tungsten cathode were considered: first, where the surface temperature of the cathode is moderate, the thermionic current density is small, and virtual cathode is not formed, and second, where the surface temperature of the cathode is extremely high and the thermionic current density is so large that the virtual cathode is formed.

In the frame of hydrodynamic 2T thermal plasma modeling, the sheath potential drop and the heat fluxes calculated by the proposed sheath model can be used in formulating boundary conditions at the wall for the electric potential and energy equations for electrons and heavy particles as in [1, 2]. Such boundary conditions enable a self-consistent calculation of electric potential distributions and heat transfers in the wall and in the arc for real arc geometries.

**References**

[1] Pekker L. and Hussary N., Effect of Boundary Conditions on the Heat Flux to the Wall in Two Temperature Modeling of Thermal Plasmas, *J. Phys. D: Appl. Phys.*, 47, 445202 (2014).

[2] Pekker L. and Hussary N., Boundary conditions at the walls with thermionic electron emission in two temperature modeling of "thermal" plasmas, *Phys. Plasmas*, 22, 083510 (2015).

[3] Godyak V. A., Modified Bohm Criterion for a Collisional Plasma, *Physics Letters*, 89A, 80-81 (1982).

[4] Godyak V. A. and Sternberg N., Smooth Plasma-Sheath Transition in a Hydrodynamic Model, *IEEE Trans. Plasma Sc.*, 18, 159-168 (1990).

[5] Bohm D., The Characteristic of Electrical Discharge in Magnetic Fields (MacGraw-Hill, New York, 1949), ch.3, p. 77 (1949).

[6] Jenista J., Heberlein J. J. R., and Pfender E, Numerical model of the anode region of High-Pressure Electric Arcs, *IEEE Trans. Plasma Sci.*, 25, 883-890 (1997).

[7] Freton P., Gonzales J. J., Ranarijaona Z., and Mougenot J., Energy equation formulations for two-temperature modeling of "thermal" plasmas, *J. Phys. D. :Appl. Phys.*, 45 465206 (2012).

[8] Capitelli M., Charge Transfer from Low-Lying Excited States: Effects on Reactive Thermal Conductivity, *J. Plasma Physics*, 14, 365-371 (1975).

[9] A. A. Phelps, Cross section and swarm coefficient for nitrogen ions in $N_2$ and argon ions in neutrals in Ar for energies from 0.1 eV to 10 kV, *J. Phys. Chem. Ref. Data*, 20, 557-573 (1990).

[10] Boulos M. I., Fauchais P., and Pfender E., Thermal Plasmas: Fundamentals and Applications (*Plenum Press, New York, N. Y.*, 1994) Vol. 1.

[11] Benilov M. S., Understanding and Modeling Plasma-Electrode Interaction in High-Pressure Arc Discharges: A Review*, J. Phys. D. :Appl. Phys.*, 41 144001 (2008).

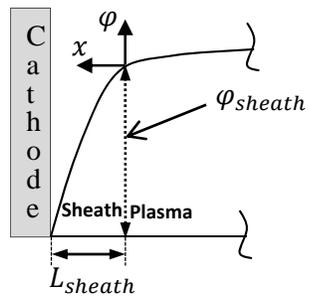

Fig. 1. Schematic of the electrical potential distribution at the cathode. $L_{sheath}$ is the sheath length and $\varphi_{sheath}$ is the sheath potential drop.

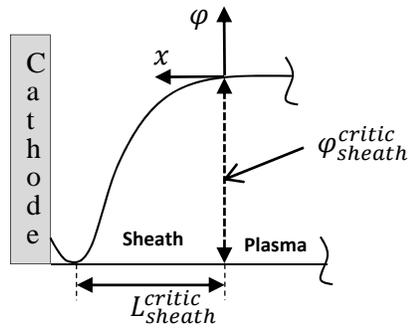

Fig. 2. Schematic of the electrical potential distribution at the cathode in the case of the virtual cathode; $L_{sheath}^{critic}$ is the critical sheath length and $\varphi_{sheath}^{critic}$ is the critical sheath potential drop corresponding to $E_{sur} = 0$.

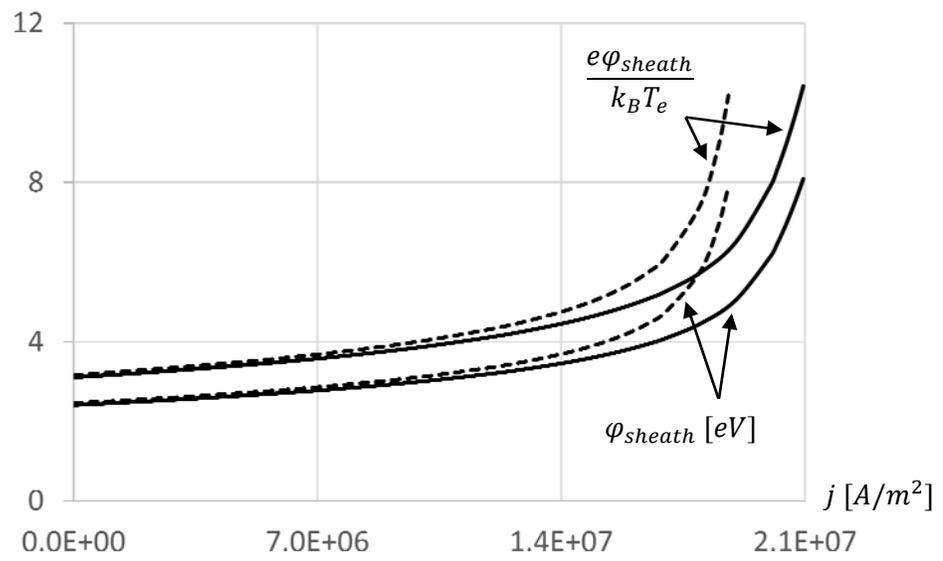

Fig. 3. The sheath models with a thermionic cathode: $\varphi_{sheath}$ and $e\varphi_{sheath}/k_B T_e$ vs. total cathode current density in the sheath; solid lines correspond to the present model and the broken lines to model [2].

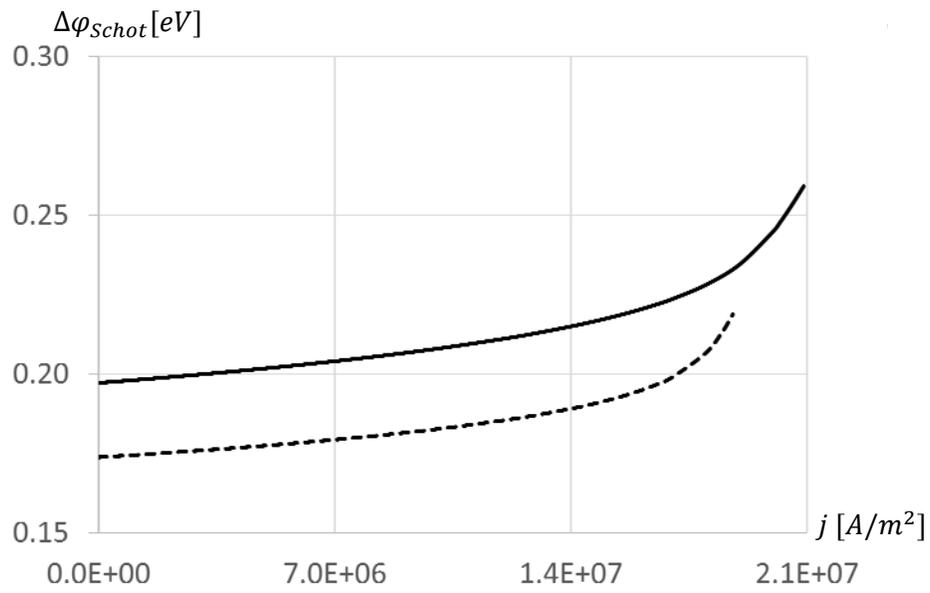

Fig. 4. The sheath models with a thermionic cathode: Schottky correction factors vs. total cathode current density in the sheath; solid line corresponds to the present model and the broken line to model [2].

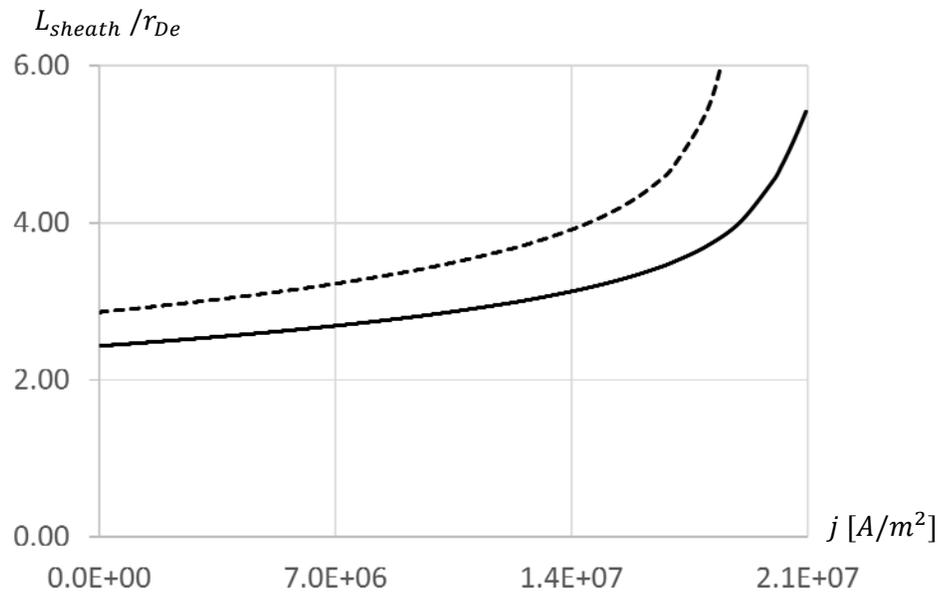

Fig. 5. The sheath models with a thermionic cathode: $L_{sheath}/r_{De}$ vs. total cathode current density in the sheath; solid line corresponds to the present model and the broken line to model [2].

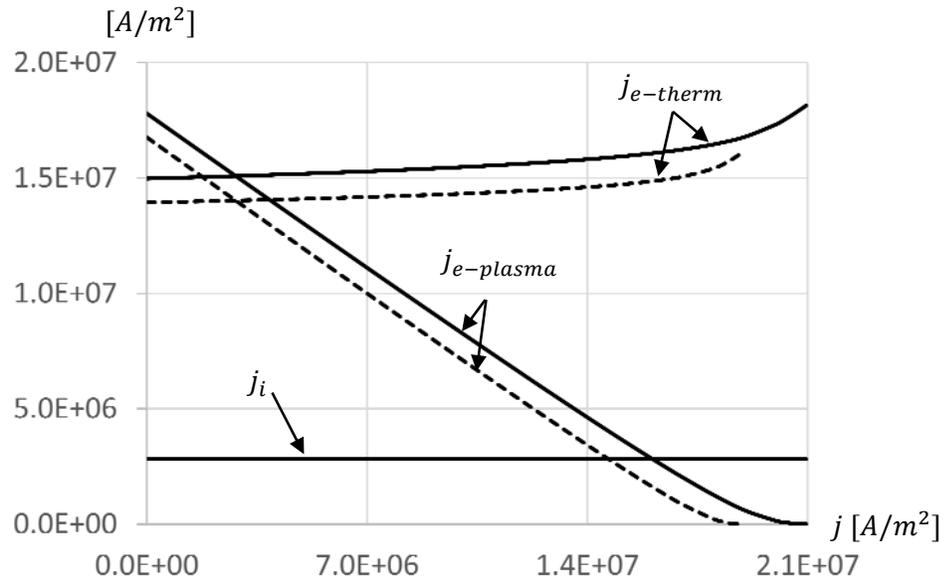

Fig. 6. The sheath models with a thermionic cathode: $j_{e-therm}$, $j_i$, and $j_{e-plasma}$ vs. total cathode current density in the sheath; solid lines corresponds to the present model and the broken lines to model [2].

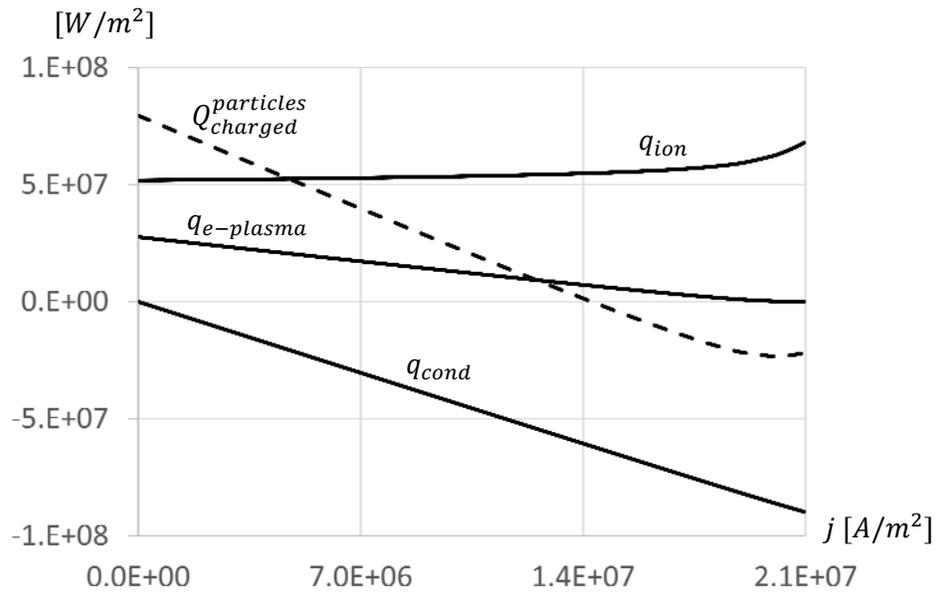

Fig. 7. The sheath model with a thermionic cathode: Solid lines – the heat fluxes to the cathode due to the charge particles, the broken line – the total heat flux to the cathode due to the charged particles, $Q_{charged}^{particles} = q_{ion} + q_{e-plasma} + q_{cond}$.

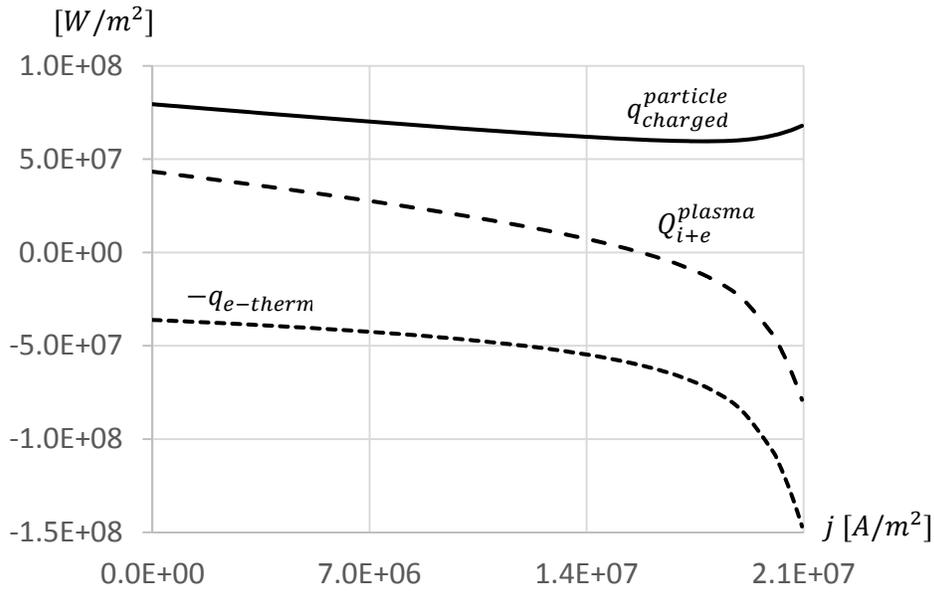

Fig. 8. The sheath model with a thermionic cathode: Solid line – the heat flux from the plasma to the cathode due to charged particles, Eq. (18); dotted line – the heat flux to the plasma due to thermionic electrons; and, the broken line – the total heat flux that the plasma loses due to charged particles, $Q_{i+e}^{plasma} = q_{charged}^{particles} - q_{e-therm}$.

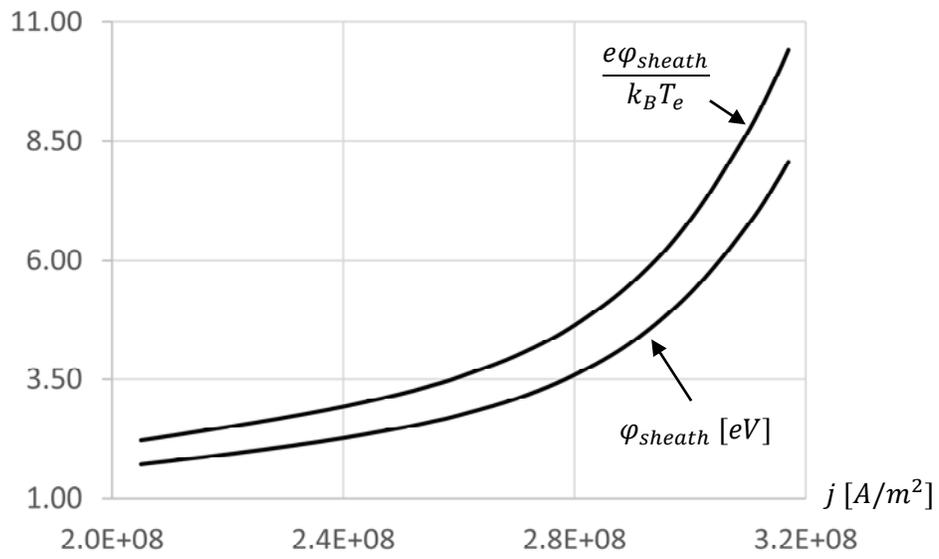

Fig. 9. The sheath model with a thermionic cathode, virtual cathode: $\varphi_{sheath}$ and $e\varphi_{sheath}/k_B T_e$ vs. total cathode current density in the sheath.

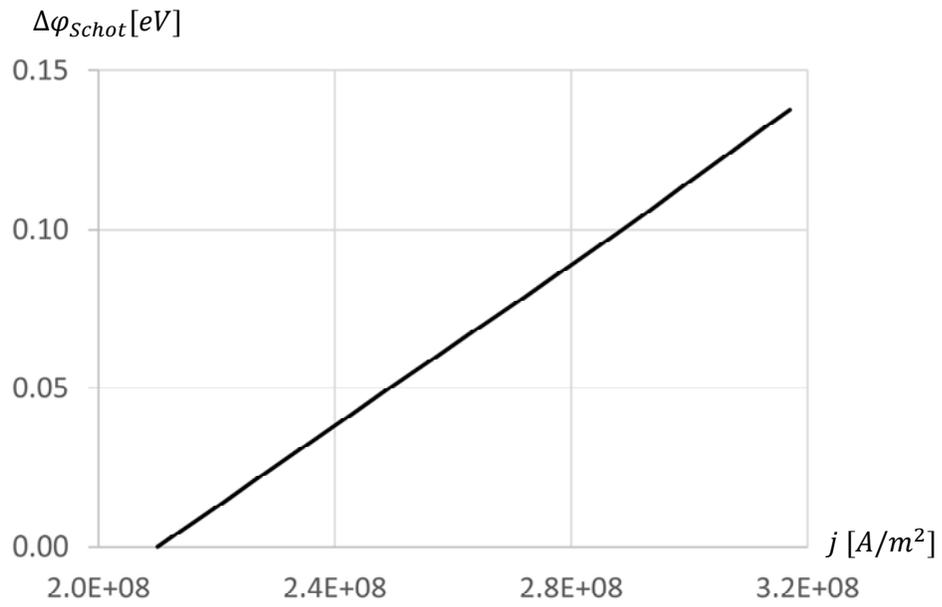

Fig. 10. The sheath model with a thermionic cathode, virtual cathode: Schottky correction factors vs. total cathode current density in the sheath.

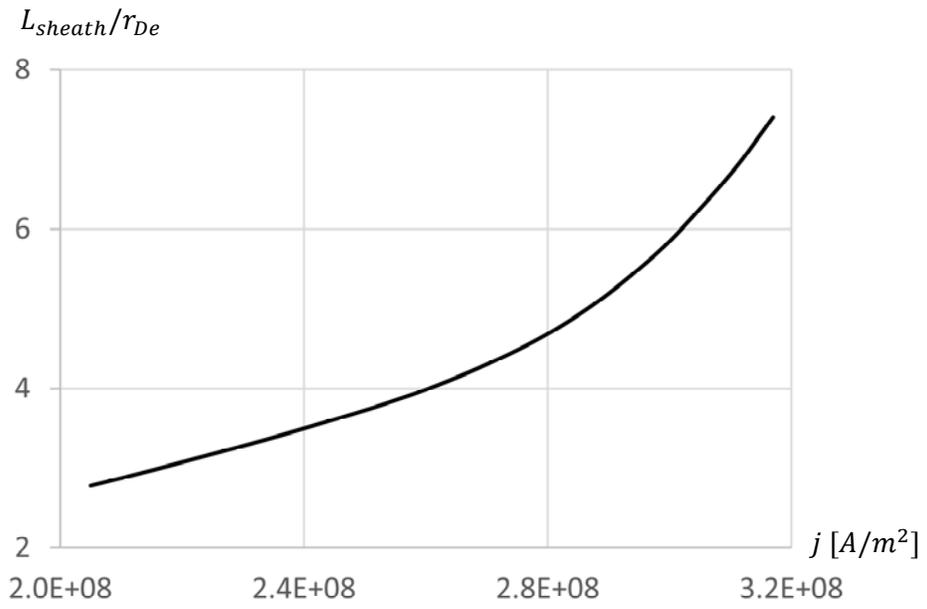

Fig. 11. The sheath model with a thermionic cathode, virtual cathode: $L_{sheath}/r_{De}$ vs. total cathode current density in the sheath.

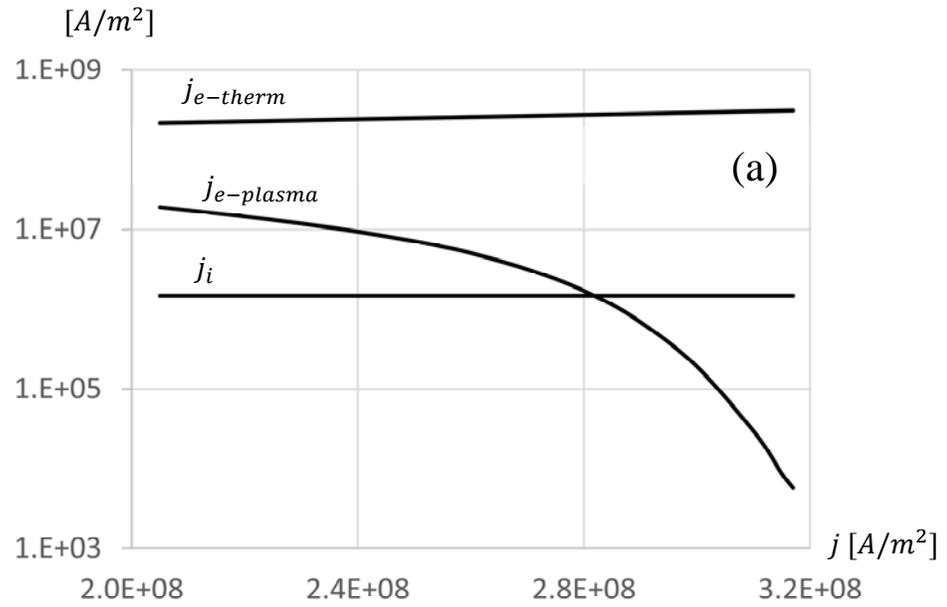

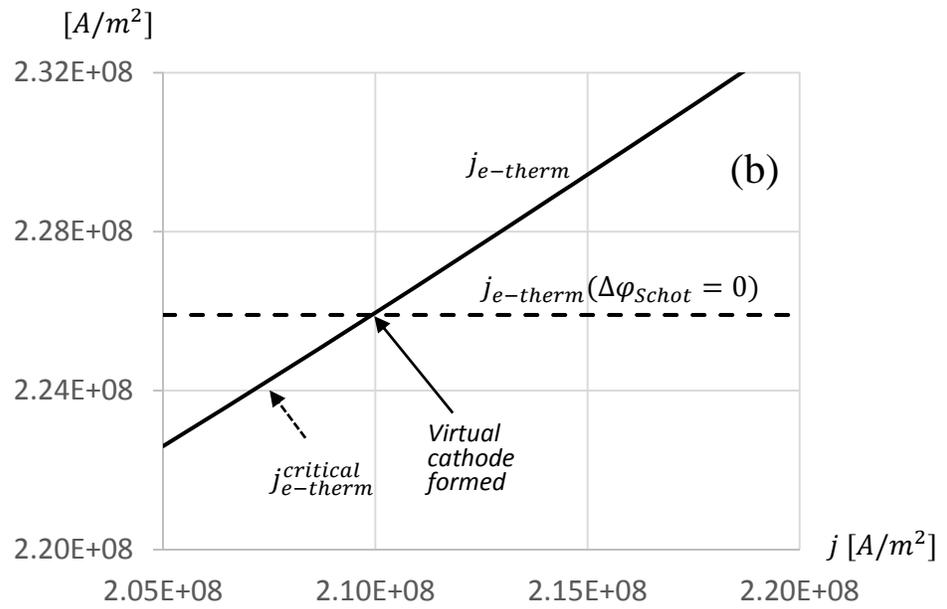

Fig. 12. The sheath model with a thermionic cathode, virtual cathode:
a - $j_{e-therm}$, $j_i$, and $j_{e-plasma}$ vs. total cathode current density in the sheath;
b - $j_{e-therm}$, and $j_{e-therm}(\Delta\varphi_{Schot} = 0)$.